\documentclass[11pt]{article}
\usepackage{graphicx}
\makeatletter

\newcommand{\singlespacing}{\let\CS=\@currsize\renewcommand{\baselinestretch}{1}\tiny\CS}
\newcommand{\oneandahalfspacing}{\let\CS=\@currsize\renewcommand{\baselinestretch}{1.25}\tiny\CS}
\newcommand{\doublespacing}{\let\CS=\@currsize\renewcommand{\baselinestretch}{1.35}\tiny\CS}

\def\@citex[#1]#2{\if@filesw\immediate\write\@auxout{\string\citation{#2}}\fi
  \def\@citea{}\@cite{\@for\@citeb:=#2\do
    {\@citea\def\@citea{,\linebreak[0]\hskip0pt plus .2em}%
      \@ifundefined{b@\@citeb}%
      {{\bf ?}\@warning{Citation `\@citeb' on page \thepage\space undefined}}%
      \hbox{\csname b@\@citeb\endcsname}}}{#1}}

\newtheorem{rule-def}[theorem]{Rule}

\begin{document}
%\setcounter{chapter}{1}
% defining short form------
\newcommand{\la}{\lambda}
\newcommand{\si}{\sigma}
\newcommand{\ol}{1-\lambda}
\newcommand{\be}{\begin{equation}}
\newcommand{\ee}{\end{equation}}
\newcommand{\bea}{\begin{eqnarray}}
\newcommand{\eea}{\end{eqnarray}}
\newcommand{\nn}{\nonumber}
\newcommand{\lb}{\label}

\noindent
{\large \bf Relativistic anisotropic charged fluid spheres with varying cosmological constant}\\

\noindent
 Saibal Ray$^1${\footnote {Permanent address: Department of Physics, Barasat Government College, Barasat 743 201, West Bengal, India}}
{\footnote {saibal@iucaa.ernet.in}}, Sumana Bhadra$^2$ and Anand
S.
Sengupta$^1$\footnote{anandss@iucaa.ernet.in}\\

\noindent
$^1${Inter-University Centre for Astronomy and Astrophysics, Post Box 4, Pune 411 007, India}\\
\noindent
$^2${ Balichak Girls' High School, Balichak, West Midnapur 721 124, West Bengal, India}\\

\noindent
{\bf Abstract.} Static spherically symmetric anisotropic source has been studied for the
Einstein-Maxwell field equations assuming the erstwhile cosmological constant
$ \Lambda $ to be a space-variable scalar, viz., $ \Lambda = \Lambda(r) $. Two cases are examined
out of which one reduces to isotropic sphere. The solutions thus obtained are shown to be electromagnetic
in origin as a particular case. It is also shown that the generally used pure charge condition,
viz., $ \rho + p_r = 0 $ is not always required for constructing electromagnetic mass models. \\

\noindent
PACS number(s): 04.20.-q, 04.20.Jb\\

\noindent
{\bf 1. Introduction}\\

\noindent The cosmological constant $\Lambda$, related to the
energy of space, introduced by Einstein in general relativity has
become of late very significant from the view point of cosmology.
Though Einstein ultimately abandoned it stating that it was a
``blunder'' in his life but Tolman keeps it as a constant
quantity in his field equations even in 1939's famous work related
to astrophysical system (Tolman, 1939). It is also, in favor of
keeping $\Lambda$, argued by Peebles and Ratra (2002) that like
all energy, the zero-point energy related to space has to
contribute to the source term in Einstein's gravitational field
equations. However, it is being gradually felt that the erstwhile
cosmological constant $\Lambda$ is indeed a scalar variable
dependent on time rather than a constant as was being believed
earlier. Recently, this variation in cosmological constant is
also observationally confirmed due to the evidence of high
redshift Type Ia supernova (Perlmutter {\it et al.}, 1998; Riess
{\it et al.}, 1998) for a small decreasing value which is $\leq
10^{-56} cm^{-2}$ at the present epoch. Obviously, once the
$\Lambda$ becomes a scalar its dependence need not be limited
only to time coordinate (as in cosmology). Since it enters in the
field equations as a variable, it must be dependent on space
coordinates as well. Therefore, in general, the $\Lambda$ is a
scalar variable dependent on, either or both, space and time
coordinates. We believe that just as in cosmology the dependence
of $\Lambda$ on time has been found to be of vital importance
playing a significant role now, its dependence on space
coordinates is equally important for astrophysical problems, in
particular, the problems of small dimensions like that of an
extended electron. With this view point, we consider here a
charged anisotropic static spherically symmetric fluid source of
physical radius, $a$, by introducing a scalar variable $\Lambda$
dependent on the radial coordinate. The field equations thus
obtained, under certain mathematical assumptions, yield a set of
solutions which has a kind of another historical importance,
known in the name of Electromagnetic Mass Models (EMMM) in the
literature (Feynman {\it et al.}, 1964). The effective
gravitational mass of these models depends on the electromagnetic
field alone (the effective gravitational mass vanishes when the
charge density vanishes). Such models have been studied by
several authors (Tiwari {\it{et al.}} 1984, 1986, 1991; Gautreau,
1985; Gr{\o}n, 1985, 1986a, 1986b; Ponce de Leon, 1987a, 1987b,
1988; Tiwari and Ray, 1991a, 1991b, 1997; Ray {\it{et al.}},
1993; Ray and Ray, 1993). All these EMMM, however, have been
obtained under a special assumption $ \rho + p = 0 $, where $
\rho $ is the matter-energy density and $ p $ is the fluid
pressure under the general condition that $\rho>0$ and $p<0$.
This type of equation of state implies that the matter
distribution is in tension and hence the matter is known, in the
literature, as a `false vacuum' or `degenerate vacuum' or
`$\rho$-vacuum' (Davies, 1984; Blome and Priester, 1984; Hogan,
1984; Kaiser and Stebbins, 1984). A natural question arises
whether there exists any EMMM where this condition is violated,
i.e., when $\rho + p \neq 0$. This is the main motivation of the
present investigation and here we have shown that even for $\rho
+ p \neq 0 $ EMMM can be constructed. However, to carry out this
plan we have assumed here that the so-called cosmological constant
is dependent on the space coordinate (Chen and Wu, 1990; Narlikar
{\it et al.}, 1991; Ray and Ray, 1993; Tiwari and Ray, 1996). This
will help us to have several
class of solutions related to EMMM. \\

\noindent
{\bf 2. Einstein-Maxwell field equations }\\
\noindent
Let us consider a spherically symmetric line element
\be
ds^{2} = g_{ij}dx^{i}dx^{j} = e^{\nu(r)} dt^{2} - e^{\lambda(r)} dr^{2} - r^{2} ( d \theta ^{2} +
sin^{2} \theta d\phi^{2} ), \qquad (i,j = 0, 1, 2, 3).
\ee
\noindent
Now, the Einstein field equations for the case of charged anisotropic source are
\be
{G^{i}}_{j} = {R^{i}}_{j} - \frac{1}{2}{{g^{i}}_{j}} R + {g^{i}}_{j}\Lambda = -\kappa [{{T^{i}}_{j}}^{(m)} + {{T^{i}}_{j}}^{(em)}] ,
\ee
where ${T^{i}}_{j}^{(m)}$ and ${T^{i}}_{j}^{(em)}$ are respectively the energy-momentum
tensor components for the anisotropic matter source and the electromagnetic
field and are given by
\be
{{T^{i}}_{j}}^{(m)} = (\rho + p_{\perp}) u^{i}u_{j} - p_{\perp} { g^{i}}_{j} +(p_{\perp} -p_{r}){\eta}^{i}{\eta}_{j}
\ee
with  $ u_{i}u^{i} = - {\eta}_{i}{\eta}^{i}=1,$ and\\
\be
{{T^{i}}_{j}}^{(em)} = \frac{1}{4\pi} [- F_{jk}F^{ik} + \frac{1}{4\pi} {g^{i}}_{j}F_{kl} F^{kl}].
\ee
The Maxwell electromagnetic field equations are given by
\be
{[{(- g)}^{1/2} F^{ij}],}_{j} = 4\pi J^{i}{(- g)}^{1/2},
\ee
and
\be
F_{[ij,k]} = 0,
\ee
where the electromagnetic field tensor $F_{ij}$ is related to the
electromagnetic potentials through $ F_{ij} = A_{i,j} - A_{j,i} $ which,
obviously, is equivalent to the equation (5), viz., $ F_{[i,j,k]} = 0 $.
Further, $u^{i}$ is the 4-velocity of a fluid element, $J^{i}$ is the
4-current satisfying $J^{i} = \sigma u^{i}$, where $\sigma$ is the
charge density, and $ \kappa = 8 \pi $ (in relativistic unit G = C = 1). Here
and in what follows a comma denotes the partial derivative with respect to the
coordinates (involving the index).\\
\noindent
The Einstein-Maxwell field equations (2)--(6) corresponding to anisotropic
charged source with cosmological variable, are then given by
\be
e^{-\lambda} ( \lambda^{\prime}/r - 1/r^{2} ) + 1/r^{2}
= 8 \pi {{T^{0}}_{0}} = 8\pi \rho + E^{2} + \Lambda  ,
\ee
\be
e^{-\lambda} ( \nu^{\prime}/r + 1/r^{2} ) - 1/r^{2}
= - 8 \pi {{T^{1}}_{1}} = 8\pi {p}_{r} -E^{2} - \Lambda,
\ee
\be
e^{-\lambda} [ \nu^{{\prime}{\prime}} + {\nu^{\prime}}^{2}/4 - {\nu^{\prime}
\lambda^{\prime}}/4 + (\nu^{\prime} - \lambda^{\prime} )/ 2r]
= - 8 \pi {{T^{2}}_{2}} = - 8 \pi {{T^{3}}_{3}}
= 8\pi p_{\perp} + E^{2} - \Lambda ,
\ee
\be
{(r^2 E)}^{\prime} = 4\pi r^2 \sigma e^{\lambda/2}.
\ee
The equation (10) can equivalently, in terms of the electric charge $q$, be expressed as
\be
q(r) = r^2E(r) = \int_{0}^{r} 4 \pi r^2 \sigma e^{\lambda/2} dr
\ee
\noindent
where $p_r$, $p_{\perp}$ and $E$ are the matter-energy density, radial and
tangential pressures and intensity of the electric field respectively. Here prime
denotes derivative with respect to radial coordinate $r$ only.\\
\noindent
The equation of continuity,  ${{{T^{i}}_{j}}};i = 0$ for the field equations
(2) - (6), is given by
\be
\frac{d}{dr}\left[p_r - (E^2 + \Lambda) /{8 \pi}\right] +
( \rho + p_r ) {\nu^\prime}/2 = E^2/2 \pi r + 2( p_{\perp} - p_r)/r.
\ee
Now, we assume the relation between the radial and tangential pressure as
\be
p_{\perp} = n p_{r},\quad(n\neq 1).
\ee
\noindent
Assuming further that the radial stress  $ {{T^{1}}_{1}} = 0 $ (Florides, 1987;
Kofinti, 1985;  Gr{\o}n, 1986b;  Ponce de Leon, 1987b; Tiwari and Ray, 1996),
one gets
\be
\nu^{\prime} = (e^{\lambda} - 1)/r,
\ee
\be
p_r = (E^2 +\Lambda)/8 \pi.
\ee
Using equations (13) - (15), in equation (12), we get
\be
\rho + p_r = [(n + 1)E^2  + (n - 1)\Lambda)]/2 \pi(e^\lambda - 1).
\ee
\noindent
Similarly, equations (7) and (8) yield,
\be
e^{-\lambda}( \nu^{\prime} + \lambda^{\prime}) = 8 \pi r( \rho + p_r).
\ee
Again, equation (7) together with equation (15), gives
\be
e^{-\lambda} = 1 - 2m(r)/r,
\ee
where m(r), called the gravitational mass, takes the form
\be
m(r) = M(r) + \mu(r) = 4 \pi \int_{0}^{r}[\rho + p_r] r^2 dr,
\ee
the Schwarzschild mass and the mass equivalence of electromagnetic field,
respectively, being defined as
\bea
M(r) = 4 \pi \int_{0}^{r} \rho r^2 dr, \quad \mu(r) = 4 \pi \int_{0}^{r} p_r r^2 dr .
\eea
\noindent
Now, from the above equation (19) it is easily observed that the condition
$ \rho + p_r = 0 $, yields a flat space-time through the equations
(18) and (14) and has been considered by us in another context (Tiwari
et al., 2000). Hence, the non-trivial solutions exits here for the case
$ \rho + p_r \neq 0 $ only.\\

\noindent
{\bf 3. Solutions for the static charged fluid spheres }\\

\noindent
Let us now solve the equation (16) under the constraint $ \rho + p_r \neq 0 $,
assuming different mathematical conditions. As the equation (16) is involved
with two physical parameters so unless we specify one parameter it is not possible
to solve the equation (19). Let us, therefore, assume the following two cases which
will yield solutions with physically interesting features as the analysis in
Section 4 demonstrates.\\
\noindent
{\bf Case I:}\\
\noindent
We now make the choice
\be
\Lambda = E^2 - N\Lambda_{0},
\ee
where $N$ is an integer and $ \Lambda_{0} $ is the erstwhile
non-zero cosmological constant.\\
With the help of equations (16) and (21), the equation (19) takes
the form \be m(r) = 2 \int_{0}^{r}[2nE^2 - (n - 1)N \Lambda_{0}]
r^2 dr/(e^{\lambda} - 1). \ee To make equation (22) integrable we
assume that \be E^2 = q^2/{r^4} = [k(e^{\lambda} - 1) (1 - R^2) +
(n - 1) N \Lambda_{0}] / 2n, \ee where $ k $ is a constant and $
R = r/a $, $ a $ being the radius of the sphere. This particular
choice for the electric intensity generates a model for charged
sphere which is physically very interesting as it is related to
EMMM as will be seen later on. Thus, the solution set is given by
\be e^{-\lambda} = 1 - AR^2 (5 - 3R^2), \ee \be e^{\nu} = (1 -
2A)^{5/4} e^{\lambda/4} exp[5B{tan^{-1}B(6R^2 - 5) - tan^{-1}B}/2
], \ee \be p_r = p_{\perp}/n = [ k(e^\lambda - 1)(1 - R^2) - N
\Lambda_{0}]/{8 \pi n}, \ee \be \rho =  [k(1 + 4n - e^\lambda)(1
- R^2) + N \Lambda_{0}]/{8 \pi n}, \ee where \be A = 4ka^2/15,
\quad B = [A/(12 - 25A)]^{1/2}. \ee \noindent Now, the exterior
field of a spherically symmetric static charged fluid
distribution described by the metric (1) is the unique
Reissner-Nordstr{\"o}m solution given by \be ds^{2} = \left[1 -
\frac{2m}{r} + \frac{q^2}{r^2}\right] dt^{2} - \left[1 -
\frac{2m}{r} + \frac{q^2}{r^2}\right]^{-1} \ dr^{2} - r^{2} ( d
\theta ^{2} + sin^{2} \theta d\phi^{2} ). \ee Then, application
of the matching condition on the boundary $ r = a $ yields the
total effective gravitational mass which is given by \be m(a) =
\frac{4}{15}ka^3 + \frac{q(a)^2}{2a}. \ee Hence, in terms of the
total gravitational mass, the total electric charge and the
radius of the sphere the constants $k$, $A$ and $B$ can be
expressed as \be k = 15(2am - q^2)/8a^4, \ee \be A = (2am -
q^2)/2a^2, \ee and \be B^2 = (2am - q^2)/[24a^2 - 25(2am - q^2)].
\ee \noindent Comparing the equation (19) (and hence via the
equation (20)) with the above equation (30) we can easily
recognize the first term ($4ka^3/15$) in the right hand side as
the total Schwarzschild mass whereas the second term ($q^2/2a$)
is the total mass equivalence of the electromagnetic field. Now,
$k$, in the first term of equation (30), is implicitly related to
the charge $q$ as is evident from the equation (23). Therefore,
in principle, the gravitational mass $m$ is
of purely electromagnetic origin.\\
\noindent
{\bf Case II:}\\
\noindent Here the choice is \be {\Lambda} =  N {\Lambda}_{0} -
E^{2}. \ee Then, from the equations (16) and (34), the
gravitational mass (equation (19)) reduces to \be m(r) = 2
\int_{0}^{r}[2E^2 + (n - 1)N \Lambda_{0}] r^2 dr/(e^{\lambda} -
1). \ee For the further assumption \be E^2 = q^2/ r^4 = [k
(e^\lambda - 1) (1 - R^2) - ( n - 1) N \Lambda_{0}]/2, \ee we
have the solution set \be e^{-\lambda} = 1 - AR^2 (5 - 3R^2), \ee
\be e^{\nu} = (1 - 2A)^{5/4} e^{\lambda/4} exp[5B{tan^{-1}B(6R^2
- 5) - tan^{-1}B}/2 ], \ee \be p = N\Lambda_{0}/8 \pi, \ee \be
\rho = [4k(1 - R^2) - N\Lambda_{0}]/8 \pi. \ee We see that $
\lambda $ and $ \nu $ retain the same form as in case I and hence
the total gravitational mass is also given by the equation (30).
Further, it is observed that the present case automatically
reduces to an isotropic one
as the pressure $p$ does not associated with the anisotropic factor $n$. \\

\begin{figure}[h]
\centering
\includegraphics[width=0.45\textwidth]{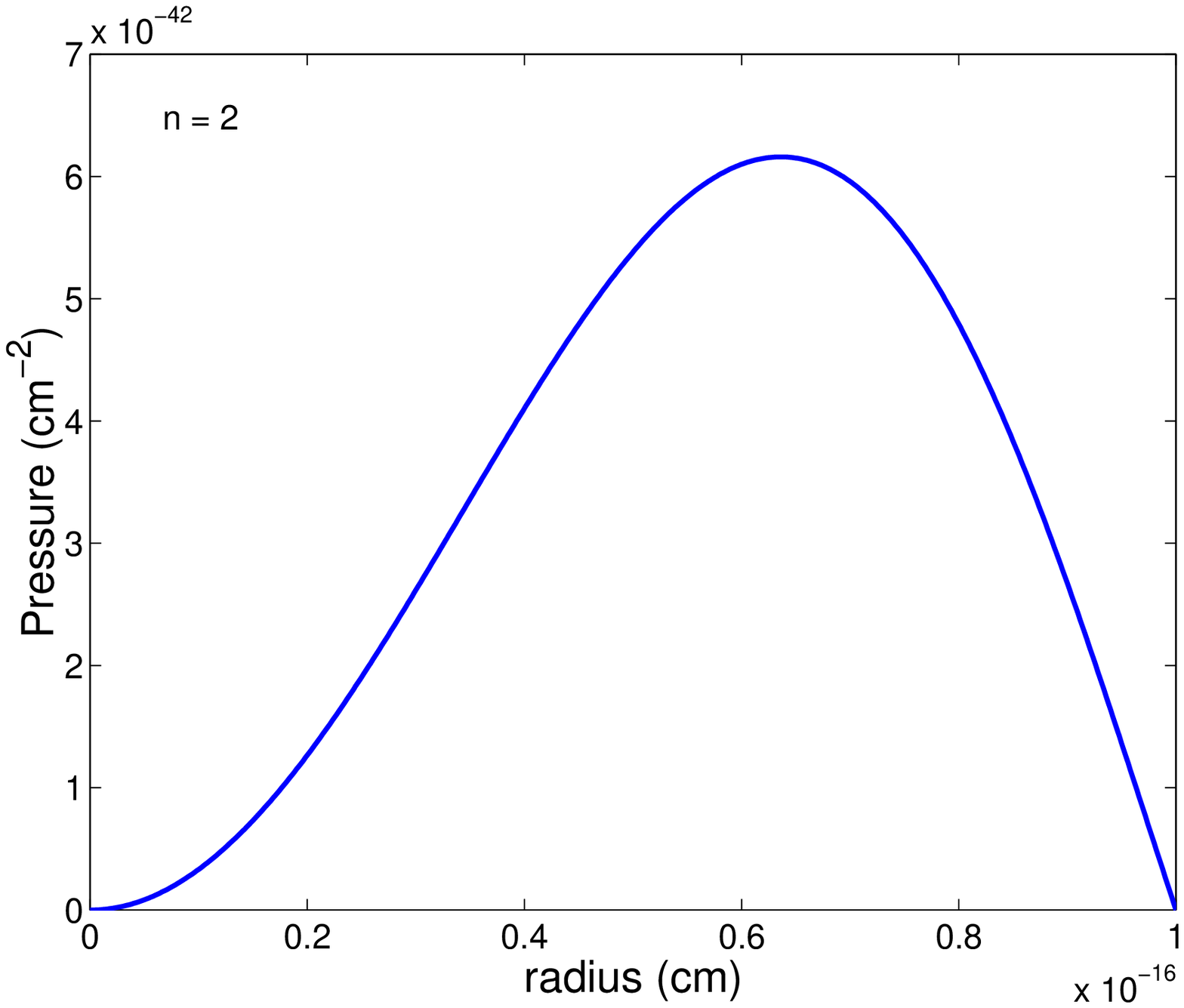}
\includegraphics[width=0.45\textwidth]{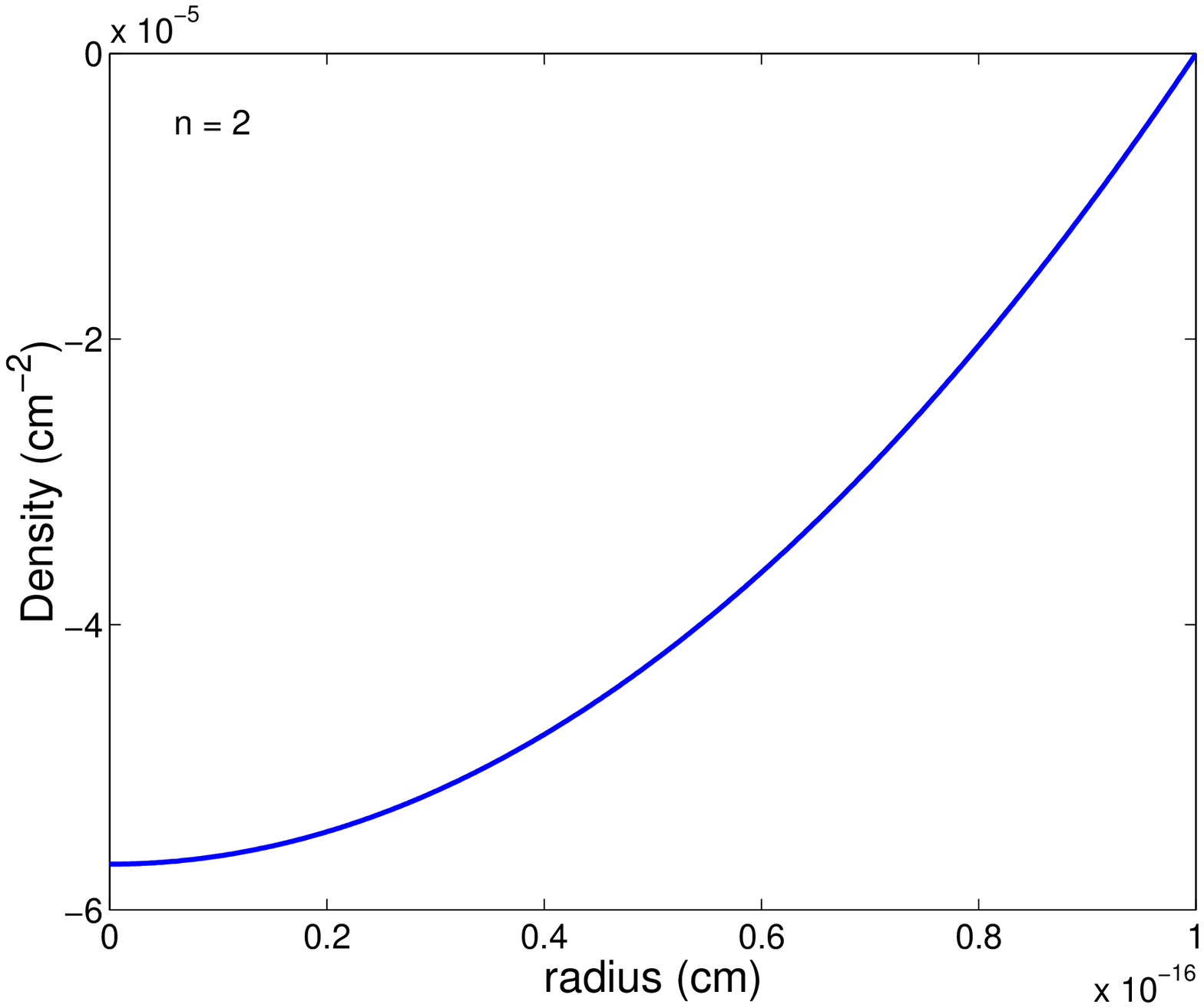}
\caption{The fluid pressure and the energy density as a function of radius
are plotted for a classical electron with radius $a = 10^{-16}$ cm,
mass $m = 6.76 \times 10^{-56}$ cm and charge $q = 1.38 \times 10^{-34}$ cm.}
\label{fig1}
\end{figure}

\begin{figure}[h]
\centering
\includegraphics[width=0.7\textwidth]{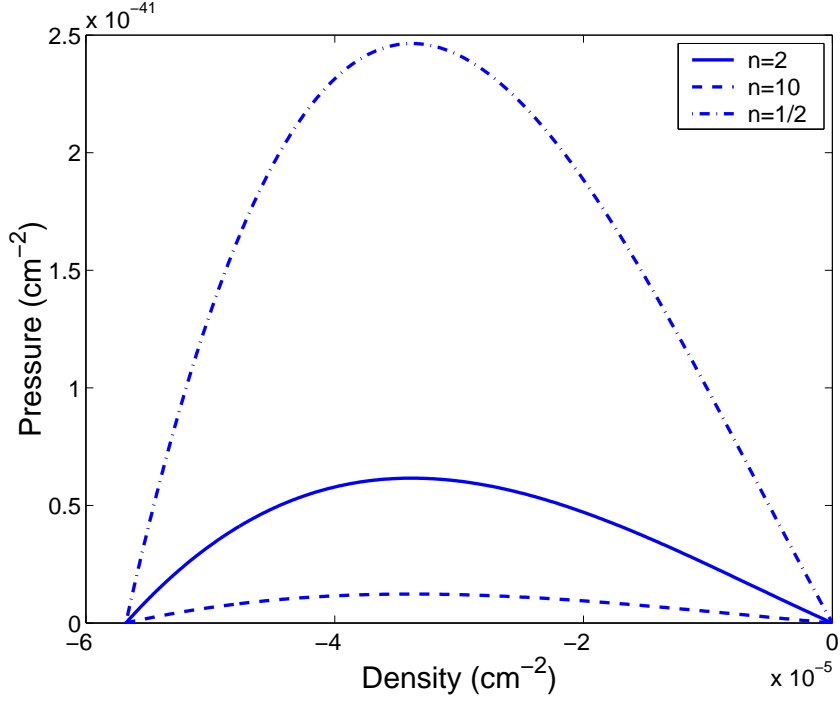}
\caption{The fluid pressure and the energy density profile is shown for
a classical electron with radius $a = 10^{-16}$ cm,
mass $m = 6.76 \times 10^{-56}$ cm and charge $q = 1.38 \times 10^{-34}$ cm.}
\label{fig2}
\end{figure}

\noindent
{\bf 4. Physical properties of the static charged fluid spheres }\\
\noindent
{\bf Case I:}\\
\noindent (1) The equation (23) indicates that for getting a
direct dependence of $k$ upon $q$ one can admit the following
relaxation such that (i) $N = 0$ when $\Lambda_0 \neq 0$ and $(n
- 1) \neq 0$, (ii) $(n - 1) = 0$ when $N \neq 0$ and $\Lambda_0
\neq 0$ and (iii) $\Lambda_0 = 0$ when $ (n - 1) \neq 0$ and $N
\neq 0$. The third possibility seems to contradict the
observational results related to the Supernova type Ia where the
cosmological constant is found to be a non-zero value whereas the
second one provides an isotropic case. Then, suitably opting for
the sub-case (i), viz., $N = 0$ one obtains \be q^2 =
k{r^4}(e^{\lambda} - 1)(1 - R^2)/2n, \ee \be p_r = p_{\perp}/n =
k(e^\lambda - 1)(1 - R^2)/{8 \pi n}, \ee \be \rho = k(1 + 4n -
e^\lambda)(1 - R^2)/{8 \pi n}. \ee \noindent Thus, for vanishing
electric charge the gravitational mass in the equation (30),
including all the physical parameters (viz., pressures and
density), vanishes
and one obtains EMMM. \\
\noindent (2) The central and the boundary pressures are found to
be equal here i.e., $p_r(0) = p_{\perp}(0)/n = - N\Lambda_0/8\pi
n$ and $p_r(0) = p_{\perp}(0)/n = - N\Lambda_0/8\pi n$
respectively whereas the respective densities are $(4nk +
N\Lambda_0)/8\pi n$ and $N\Lambda_0/8\pi n$. So, the present
model has a constant pressure throughout the sphere though the
density decreases from center to boundary. For the value $N = 0$,
however, we have zero pressure, both at the center and boundary,
and the density decreases from the central value $k/2\pi$ to zero
at the boundary. Thus, with $N = 0$ the present model goes to a
physically well-behaved static charged dust case. However, we
would like to investigate how the pressure and the density behave
in between the center to boundary. To see this, as an example, we
have plotted the graph for the extended classical electron of
Lorentz type (1904) in Figure \ref{fig1}. It shows the nature of
variation of the fluid pressure and the energy density with
radius from the center of the matter distribution to the
boundary. We observe that though the pressure is zero at the
center as well as at the boundary but the in-between feature is
very interesting. Starting from zero it increasing slowly and
then after attaining a maximum value drops steadily to zero. In
the case of density the curve starts from the non-zero central
value $k/2\pi$ and then smoothly decreases to zero at the
boundary. However, it is to be noted here that for the electron
with radius $a = 10^{-16}$ cm, mass $m = 6.76 \times 10^{-56}$ cm
and charge $q = 1.38 \times 10^{-34}$ the value of the constant
$k$ becomes a negative quantity with $- 3.568215 \times 10^{-4}$.
This particular aspect one should keep in mind for the
explanation of the internal structure of classical electron with
respect to the fluid pressure and the energy density. Thus we see that
the density starts from the down of the axis due to this negative value
attached to it. In the Figure \ref{fig2} the pressure-density profile
is shown for different values of the anisotropic factor $n$.\\

\noindent (3) In the above analysis for the general value of $N$
we have seen that the fluid pressure is altogether negative
whereas the density is a positive quantity. Now, it can be
observed from the equations (26) and (27) that $ \rho + p_r \neq
0 $ except at the boundary where it is equal to $ p_r(a) = -
\rho(a) $, where $\rho(a) > 0$ and $p(a) < 0$ as seen earlier.
Otherwise, it will have a general value $ k (1 - R^2)/2\pi$ . The
central value is then $ p_r(0) = - \tilde{\rho}(0) $ where
$\tilde{\rho}(0) = \rho(0) -  k/2\pi$. As the model demands for
$\rho > 0$ and $p < 0$ so the condition to be satisfied here is
$\rho(0) >  k/2\pi$. The general condition for the negative
pressure and positive density is then $\rho >  k (1 - R^2)/2\pi$
for all $r \leq a$. These results are also true
for the sub-case $N = 0$.\\

\begin{figure}[h]
\centering
\includegraphics[width=0.6\textwidth]{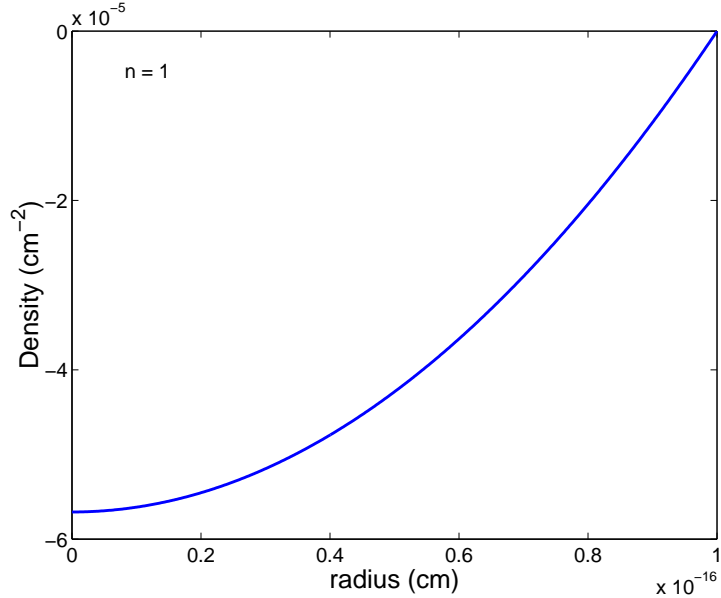}
\caption{The energy density as a function of radius is plotted for a classical electron
with radius $a = 10^{-16}$ cm, mass $m = 6.76 \times 10^{-56}$ cm and charge
$q = 1.38 \times 10^{-34}$.}
\label{fig3}
\end{figure}

\noindent
{\bf Case II:}\\
\noindent (1) Here for $N = 0$ the solution set regarding the
pressure and density (vide equations (39) and (40)) reduces to \be
p = 0, \ee \be \rho = k(1 - R^2)/2 \pi, \ee when the electric
charge is given by \be q^2 = k (e^\lambda - 1) (1 - R^2)r^4/2.
\ee Thus, as in the previous case, for $q = 0$ we get $k = 0$
which in turn makes mass and density everything zero and also the
space-time becomes flat.
Thus, the model presented here is an EMMM.\\
\noindent (2) In the present case also the central and the
boundary pressures are equal with a value $N\Lambda_0/8\pi$ and
the respective densities are $(4k - N\Lambda_0)/8\pi $ and $ -
N\Lambda_0/8\pi$. The pressures become zero for $N = 0$,
 both at the center and boundary, and densities have the positive central value $k/2\pi$
whereas the boundary-value is zero. Thus, we again get a
physically interesting charged dust case with $N = 0$, only
difference with the previous Case I is that this is now a case of
isotropic fluid sphere. Figure \ref{fig3} shows the nature of
variation of energy density with radius from the center of the
matter distribution to the boundary for an extended classical
electron. Here also the behavior is regular and well defined.\\

\noindent (3) In the present case also, by virtue of equations
(39) and (40), $\rho + p_r \neq 0 $ which reads here as $\rho +
p_r =  k (1 - R^2)/2\pi$. Here the central value, at $r=0$, is $
\rho = - \tilde{p}$ where $\tilde{p} = ( p - k/2\pi)$ and the
boundary one is $ \rho = - p $. Due to negativity of the density
here the condition on the pressure to be imposed is $ p >
k/2\pi$. Thus, the present Case II clearly provides an EMMM even
with a positive pressure and therefore contradicts the comment
made by Ivanov (2002) that ``... electromagnetic mass models all
seem to have negative pressure.'' The same result, i.e. the
positivity of pressures are also available in some cases of the
work done by Ray and Das (2002) related to EMMM. However, the
explanation given here is valid for any positive value of $N$ and
so the situation may completely be opposite with any negative
value of $N$. At this stage, we feel, we should not put any
restriction on the choice of the value of $N$. This is because,
in general, for a fluid sphere we should have $p \geq 0$ and
$\rho \geq 0$ so that the weak energy conditions are satisfied.
But there are also some special situations available within the
spherical system (particularly, in the case of electron with the
radius $\sim 10^{-16}$ cm) where the energy condition is violated
due to negative energy density (Cooperstock and Rosen, 1989;
Bonnor and Cooperstock, 1989). Thus, choosing the proper signatures
of $N$ we can have a class of models with diverse characters.  \\

\noindent
{\bf 5. Role of $\Lambda$: Previous
 and Present Status}\\

\noindent
The cosmological constant was introduced by Einstein in his field equation to
obtain a static cosmological solution because of the fact that due to
gravitational pull everything will collapse to a point and hence a un-wanting
situation of singularity will take place. However, he was not satisfied with
this new physical quantity as it seemed to violate Machian principle which he
tried to incorporate in the framework of his general theory of relativity. He
thus, ultimately rejected it mainly for two reasons: (i) that the
theoretical work of de Sitter showing that the Einstein's field equations
admitted a solution for empty Universe and (ii) that the experimental
discovery of expanding Universe by Hubble.\\
\noindent As stated in the introduction, the concept of
cosmological constant has been revived recently in the case of
early Universe scenario and even in particle physics. It is
gradually being felt that $\Lambda$, the erstwhile cosmological
constant is available rather than a constant, as was being
believed earlier, varying with space or time or both (Ray and
Ray, 1993; Tiwari and Ray, 1996; Tiwari et al., 2000). Further,
$\Lambda$ may be positive or negative (by imposing the condition
that its value is not equal to zero). For instance, according to
Zel'dovich (1968) the effective gravitational mass density of the
polarized vacuum is negative. Similarly, the equation of state
$\rho + p = 0$, employed by Tiwari et al. (1984) to construct
EMMM as a solution of Einstein-Maxwell field equations, provides
negative pressure. It may be emphasized here that positive
density has significant, rather major role in inflationary
cosmology whereas negative density has influence on elementary
particle models. The gravitational mass inside the spherical
charged body is negative for $r<5a/4$, where $r$ is the radial
coordinate and $a$ is the radius of the sphere (Gautreau, 1985).
It is argued by Gr{\o}n (1986a, b) that this negative mass and the
associated gravitational repulsion is due to the strain of the
vacuum because of vacuum polarization. He also argue that if a
vacuum has a vanishing energy, then its gravitational mass will be
negative and the observed expansion of the universe may be
explained as a result of repulsive gravitation. Now, if we
consider a negative $ \Lambda$ having a repulsive nature as was
considered by Einstein then this gets the same status of negative
pressure and also can be identified with the Poincare' stress.
This repulsive gravitation associated with negative $\Lambda$ can
also be explained as the source of gravitational blue shift
(Gr{\o}n, 1986a). On the contrary, positive $\Lambda$ will be
related to gravitational red shift. It may be also pointed out
that according to Ipser and Sikivie (1984) domain walls are
sources of repulsive gravitation and a spherical domain wall will
collapse. To overcome this situation the charged ``bubbles" with
negative mass keep the wall static and hence in equilibrium. In
this regard, we may also add that $\Lambda$, via repulsive
gravitation, is related to domain walls and playing an important
physical role.\\
\noindent
Very recent observations conducted by the SCP and HZT (Perlmutter {\it et al.}, 1998;
Riess {\it et al.}, 1998; Filippenko, 2001; Kastor and Traschen, 2002) show that the present
value of $\Lambda$ is positive one and hence related to the repulsive pressure.
It is believed that the present state of acceleration dominated universe is
due to the driven force of this $\Lambda$. It is to be noted that the negative $\Lambda$
corresponds to a collapsing situation of the universe (Cardenas {\it et al.}, 2002).\\

\noindent
{\bf 6. Conclusions}\\

\noindent
(i) In both the above cases I and II, it is possible to show that EMMM also can be
obtained, in principle, using the constraint $ \rho + p_r \neq 0 $. This
particular point remained unnoticed by Gr{\o}n (1986a, b) and Ponce de
Leon (1987a, b), both. However, as $\rho + p_r = 0 $ is related to vacuum polarization,
black  hole physics and inflationary cosmology (Gr{\o}n, 1985, 1986a; Wenda and
Shitong, 1985a, b; Guth, 1981; Linde, 1984), so the present model
, in general, is in contradiction to those phenomenological explanations.\\
(ii) It can be noted that in terms of energy-momentum tensor of
the fluid the condition $ \rho + p_r = 0 $ implies $ {T^{1}}_{1}
= {T^{0}}_{0}$ (Herrera and Varela, 1996) whereas $ \rho + p_r
\neq 0 $ constraint may be expressed as $ { T^{1}}_{1} = 0 $ as
we have adopted in the present approach. It is also interesting
to note that $ \rho + p_r = 0 $ and hence $ {T^{1}}_{1} =
{T^{0}}_{0}$ can be expressed in terms of the metric tensors
(vide equation (1)) as $g_{00}g_{11} = -1$. A
coordinate-independent statement of this relation is obtained by
Tiwari {\it et al.} (1984) by using the eigen values of the
Einstein tensor ${G^{i}}_{j}$.\\
(iii) Another point is that in both the cases we have taken the
assumptions in a way so that the cosmological variable $ \Lambda
$ does not vanish rather may be at most equal to $ \Lambda_{0} $,
the erstwhile cosmological constant having a finite non-zero
value. Without considering $ \Lambda_{0} $ we will have $ \Lambda
= 0 $ at the boundary $ r = a $ which is a bit unphysical and may
create difficulties, such as to entropy like problems (Beesham,
1993). In this regard, one can observe that the solutions
obtained by Gr{\o}n(1986a, b) and Ponce de Leon (1987a, b)
represent a neutral system, viz., though the net charge is not
zero but the charge on the surface of the spherical system
vanishes. The models of the present paper, in general, do not
correspond to this situation because of the fact that the
cosmological parameter $\Lambda$ does not vanish at the boundary,
rather, its value is $\Lambda_{0}$ at $r = a$. Therefore, the
present solutions correspond  to charged sphere. Of course, for
$N = 0$, like Gr{\o}n (1986, b)  and Ponce de Leon (1987a, b), we
have neutral spheres (vide equation (41) of case I and equation
(46) of case II). Thus, we have a class of solutions related to
charged as well as neutral systems depending
on the values of $N$.\\
(iv) We have, as a special case, applied the present models to
the classical electron of Lorentz type. However, it seems that
 a more realistic application of the models with variable
 cosmological constant is possible in the case of charged massive
 astrophysical systems, like neutron stars. This aspect will be
 carried out in detail in the future investigations.\\

\noindent
{\bf Acknowledgment}\\
One of the authors (SR) thanks IUCAA, Pune for providing Associateship Programme under which a part of this
work was carried out. The authors are grateful to Prof. G. Mohanty,
Sambalpur University for his critical comments which made it possible
to improve the paper. A. S. Sengupta would like to thank CSIR (India) for Senior Research Fellowship.\\

\noindent
{\bf References}\\

\noindent
Beesham A K 1993 {\it {Phys. Rev. D}} {\bf {48}} 3539\\
\noindent
Blome J J and Priester W 1984 {\it Naturwissenshaften} {\bf 71} 528\\
\noindent
Bonnor W B and Cooperstock F I 1989 {\it {Phys. Lett. A}} {\bf {139}} 442\\
\noindent
Cardenas R {\it et al.} 2002 {\it Preprint} astro-ph/0206315\\
\noindent
Chen W and Wu Y S 1990 {\it{Phys. Rev. D}} {\bf{41}} 695\\
\noindent
Cooperstock F I and Rosen N 1989 {\it{Int. J. Theor. Phys. }} {\bf{28}} 423\\
\noindent
Davies C W 1984 {\it Phys. Rev.} {\bf D30} 737\\
\noindent
Feynman R P, Leighton R R and Sands M 1964 {\it{The Feynman
   Lectures on Physics}}(Addison-Wesley, Palo Alto, Vol. II, Chap. 28)\\
\noindent
Filippenko A V 2001 {\it Preprint} astro-ph/0109399\\
\noindent
Florides P S 1997 {\it{Proc. R. Soc. (London) Ser. A}} {\bf{337}} 529\\
\noindent
Gautreau R 1985 {\it{Phys. Rev. D}} {\bf{31}} 1860\\
\noindent
Gr{\o}n {\O} 1985 {\it{Phys. Rev. D}} {\bf{31}} 2129\\
\noindent
Gr{\o}n {\O} 1986a {\it{Am. J. Phys.}} {\bf{54}} 46\\
\noindent
Gr{\o}n {\O} 1986b {\it{Gen. Rel. Grav.}} {\bf{18}} 591\\
\noindent
Guth A 1981 {\it {Phys. Rev. D}} {\bf {23}} 347\\
\noindent
Herrera L and Varela V 1996 {\it {Gen. Rel. Grav.}} {\bf {28}} 663 \\
\noindent
Hogan C 1984 {\it Nature } {\bf 310} 365\\
\noindent
Ipser J and Sikivie P 1984 {\it{Phys. Rev. D}} {\bf{30}} 712\\
\noindent
Ivanov B V 2002 {\it{Phys. Rev. D}} {\bf{65}} 104001\\
\noindent
Kaiser N and Stebbins A 1984 {\it Nature } {\bf 310} 391\\
\noindent
Kastor D and Traschen J 2002 {\it Class. Quantum Grav.} {\bf 19} 5901\\
\noindent
Kofinti N K 1985 {\it {Gen. Rel. Grav.}} {\bf {17}} 245\\
\noindent
Linde A D 1984 {\it {Rep. Prog. Phys. }} {\bf {47}} 925\\
\noindent
Lorentz H A 1904 {\it {Proc. Acad. Sci., Amsterdam}} {\bf {6}} (Reprinted in Einstein {\it et al.}, {\it The Principle of Relativity}, Dover, INC, 1952, p. 24)\\
\noindent
Narlikar J V,  Pecker J -C and Vigier J -P 1991 {\it J. Astrophys. Astr.} {\bf 12} 7\\
\noindent
Peebles P J E and Ratra B 2002 {\it Preprint} astro-ph/0207347\\
\noindent
Perlmutter S {\it et al.} 1998 {\it Nature} {\bf 391} 51\\
\noindent
Ponce de Leon J 1987a {\it{J. Math. Phys.}} {\bf{28}} 410\\
\noindent
Ponce de Leon J 1987b {\it{Gen. Rel. Grav.}} {\bf{19}} 797\\
\noindent
Ponce de Leon J 1988 {\it{J. Maths. Phys.}} {\bf{29}} 197\\
\noindent
Ray S and Das B 2002 {\it{Astrophys. Space Sci.}} {\bf{282}} 635\\
\noindent
Ray S, Ray D and Tiwari R N 1993 {\it{Astrophys. Space Sci.}} {\bf{199}} 333\\
\noindent
Ray S and Ray D 1993 {\it{Astrophys. Space Sci.}} {\bf{203}} 211\\
\noindent
Riess A G {\it et al.} 1998 {\it Astron. J.} {\bf 116} 1009\\
\noindent
Tiwari R N, Rao J R and Kanakamedala K K 1984 {\it{Phys. Rev. D}} {\bf{30}} 489\\
\noindent
Tiwari R N, Rao J R and Kanakamedala K K 1986, {\it{Phys. Rev. D}} {\bf{34}} 1205\\
\noindent
Tiwari R N, Rao J R and Ray S 1991 {\it{Astrophys. Space Sci.}} {\bf{178}} 119\\
\noindent
Tiwari R N and Ray S 1991a {\it{Astrophys. Space Sci.}} {\bf{180}} 143\\
\noindent
Tiwari R N and Ray S 1991b {\it{Astrophys. Space Sci.}} {\bf{182}} 105\\
\noindent
Tiwari R N and Ray S 1996 {\it{Ind. J. Pure Appl. Math.}} {\bf{27}} 907\\
\noindent
Tiwari R N and Ray S 1997 {\it{Gen. Rel. Grav.}} {\bf{29}} 683\\
\noindent
Tiwari R N Ray S and Bhadra S 2000 {\it{Ind. J. Pure Appl. Math.}} {\bf{31}} 1017\\
\noindent
Tolman R C 1939 {\it Phys. Rev.} {\bf 55} 367\\
\noindent
Wenda S and Shitong Z 1985a {\it{Gen. Rel. Grav.}} {\bf {17}} 739\\
\noindent
Wenda S and Shitong Z 1985b {\it {Nuo. Cim.}} {\bf{85B}} 142\\
\noindent
Zel'dovich Ya B 1968 {\it{Sov. Phys.}} {\bf{11}} 381\\

\end{document}